# Statistical Quality and Reproducibility of Pseudorandom Number Generators in Machine Learning technologies

**Benjamin Antunes**
Department of Science, University of Perpignan Via Domitia (UPVD), France

**Abstract**

Machine learning (ML) frameworks rely heavily on pseudorandom number generators (PRNGs) for tasks such as data shuffling, weight initialization, dropout, and optimization. Yet, the statistical quality and reproducibility of these generators—particularly when integrated into frameworks like PyTorch, TensorFlow, and NumPy—are underexplored. In this paper, we compare the statistical quality of PRNGs used in ML frameworks (Mersenne Twister, PCG, and Philox) against their original C implementations. Using the rigorous TestU01 BigCrush test suite, we evaluate 896 independent random streams for each generator. Our findings challenge claims of statistical robustness, revealing that even generators labeled "crush-resistant" (e.g., PCG, Philox) may fail certain statistical tests. Surprisingly, we can observe some differences in failure profiles between the native and framework-integrated versions of the same algorithm, highlighting some implementation differences that may exist.

Keywords: PRNGs, Machine learning, Reproducibility

## 1. Introduction

Modern machine learning (ML) research is heavily reliant on high-level programming languages and development frameworks. Python stands as the dominant language in this field, leading to the widespread integration of tools such as TensorFlow, PyTorch, and NumPy. This paper explores the statistical integrity of pseudorandom number generators (PRNGs) used in these environments and contrasts their statistical quality with that of their original reference implementations. Despite the critical role of PRNGs, there is limited research analyzing their statistical behavior within ML platforms. Python's built-in PRNG employs the Mersenne Twister (MT) [1]; PyTorch also uses MT, while TensorFlow defaults to Philox (with Threefry, another member of its cryptographic family, available) [2]. NumPy, in contrast, supports a broader array of generators and defaults to PCG [3]. In our study, we evaluate the statistical consistency and reproducibility of three main PRNGs—MT, Philox, and PCG—against their C reference implementations. As detailed by Salmon et al. [2], the Philox and Threefry algorithms, presented in 2011, utilize cryptographic primitives reminiscent of AES, offering high statistical quality at the cost of speed, but described as more suitable for GPU computation. For our evaluations, we rely on TestU01 [4], a comprehensive PRNG testing suite offering over 100 assessments under its "Big Crush" test battery. PCG, introduced by O'Neill in 2014, claims superior statistical quality, though this has yet to be fully validated through extensive TestU01 trials. Originally released in 1998 by Matsumoto and Nishimura, MT is recognized for its extremely long period and was enhanced in 2002 for improved initialization. The 2006 SFMT version, by Saito and Matsumoto [5], further capitalizes on modern CPU architecture, offering higher speed and better statistical properties. A GPU-friendly version, MTGP, also exists. However, the MT family is unsuitable for cryptographic tasks, and while some statistical weaknesses are known, their impact in practice appears minimal, as it remains prevalent in scientific libraries.

Evaluating PRNGs involves statistical analysis to differentiate strong generators from weak ones. Early methodologies stem from Donald Knuth's foundational work in "The Art of Computer Programming", which still holds relevance. In 1996, Marsaglia created Die Hard, a battery of 15 statistical tests, though its original code is only available through archived sources. Brown and collaborators in Australia extended this suite into Die Harder, available as open-source software. The National Institute of Standards and Technology (NIST) developed the Statistical Test Suite (STS), especially suited for cryptographic PRNG validation [6]. TestU01, designed by L'Ecuyer and Simard, stands out by offering detailed and hierarchical PRNG evaluations through its Small Crush,

Crush, and Big Crush batteries. For our tests, we employed Big Crush, which includes 106 rigorous statistical tests.

Random sampling plays a fundamental role in training AI models. For example, in "General Game Playing," (GGP) where agents must adapt to unfamiliar games based solely on rules, annual competitions hosted by Stanford highlight the value of stochastic strategies. Since 2007, Monte Carlo Tree Search (MCTS) has emerged as a dominant approach in Go and has become standard in top GGP agents and even some bioinformatics applications [7]. A critical scientific concern is reproducibility [8], essential for confirming, extending, or refuting results. In ML, reproducibility ensures that the same outcomes are obtained across repeated runs, assuming consistent settings. This is vital for debugging, comparative evaluation, and reliability. PRNGs are central to this, as many operations—like data partitioning and parameter initialization—depend on them. Without deterministic PRNG behavior, even slight variations can compound into significant performance differences. On a broader scale, reproducibility underpins scientific trust and progress [9]. If findings can be reliably replicated, they serve as a sturdy base for ongoing research. A comprehensive review of reproducibility issues across computing is available in Computer Science Review [10]. In this study, we investigate whether ML libraries and frameworks produce pseudorandom numbers of equal statistical quality to those generated by the original PRNG implementations.

We begin with a review of related work on randomness in ML, describe our experimental approach, present comparative results, and conclude with implications and suggestions for future work.

## 2. Related work

Highlighting the influence of PRNG statistical quality on neural network training, Huk [11] recently investigated how changing the PRNG can affect CNN and MLP classification results. They computed 95% confidence intervals for several quality metrics across various PRNGs. The findings revealed slight differences, indicated by distinct confidence intervals, suggesting that the choice of PRNG might significantly influence training quality, possibly requiring doubled evaluation confidence intervals. Koivu et al. [12] similarly identified a link between PRNG statistical behavior and the dropout method's impact on neural networks. Further work is essential to examine different architectures, understand PRNG quality implications, and replicate findings, given the limited research in this area. The influence of PRNGs in ML is understudied and deserves deeper investigation, especially as stochastic components grow more critical in modern ML for their computational advantages. PRNGs are integral in many aspects of ML systems. For instance, stochastic gradient descent (SGD) is a fundamental optimization method, updating parameters with single or mini-batch samples instead of the full dataset. Lu et al. [13] employed a quasi-Monte Carlo approach with a fixed scan order to accelerate learning from augmented data. Beyond SGD, regularization techniques like dropout depend heavily on randomness to function effectively. Dropout helps counter overfitting—when a model fits training data too closely—by randomly disabling neurons and their connections during training. Another randomness-based technique, stochastic depth, simplifies training deep networks by randomly bypassing layers using identity mappings, which can reduce time and enhance performance [14]. Randomness also enhances data augmentation strategies, which increase training set size through altered or synthetic data. For image tasks, augmentation might involve rotations, flips, or crops. Algorithms such as Expectation-Maximization and Markov chain Monte Carlo sampling also leverage randomness [15]. In deep learning, random augmentations range from geometric and color transformations to advanced methods like image mixing, erasing, and neural style transfer. Test-time augmentation, which varies evaluation data, further strengthens model generalization [16]. Bootstrapping, which uses random sampling with replacement to create varied datasets, reinforces these strategies. This method enhances stability and accuracy, especially in ensemble models, by exposing them to broader data variation [17]. Recent surveys note that randomness in ML helps balance efficiency and performance demands [18]. PRNGs are found across numerous ML applications: Bayesian neural networks [19], variational autoencoders [20], and reinforcement learning [21], to name a few. Techniques such as gradient noise injection also use randomness to improve training [22]. Recent studies address energy-efficient pseudorandom generation in neural systems. For instance, Kim et al. [23] applied stochastic computing (SC) in deep networks to boost latency and power metrics. SC, pioneered by John Von Neumann in the 1960s, uses random bit streams and bitwise operations to execute complex functions. Yet, Liu et al. [24] caution that SC may still be energy-intensive in deep learning contexts. The machine learning field has also embraced Transformer-based architectures, now common across domains. While GANs have found success in image synthesis [25], Transformers are best known for powering large language models. These models still depend on randomness in their training processes—mainly via dropout and SGD—to prevent overfitting and support model robustness. Pranav et al. [26] emphasize that the integrity of the random number generator is crucial, as attackers could exploit weak randomness to undermine ML models. Moreover, pseudorandom generation is a foundational concept in computational learning theory, playing a role in Probably Approximately Correct (PAC) learning [27]. Real-world applications, such as predicting droplet merging in microfluidic systems [28, 29], utilize random forest algorithms that inherently depend on PRNGs. While abstraction layers may hide these dependencies, they significantly shape algorithm performance and behavior. Gundersen et al. [30] enumerate multiple reproducibility challenges in ML, among which poorly managed PRNGs are a key factor. This brief review makes it clear: randomness and PRNGs are foundational to AI's growth and practical deployment. Despite their pervasive presence, the quality of PRNGs within ML frameworks remains insufficiently explored. Our research aims to address this gap.

## 3. Materials and methods

To address the questions raised in the introduction, we selected widely adopted ML frameworks—PyTorch and TensorFlow—along with the NumPy library, due to their prevalence and significance in the ML ecosystem. We conducted a comprehensive statistical analysis of the native implementations of the PCG and Philox PRNGs, as these are commonly integrated into ML technologies. The latest version of TensorFlow recommends using a Generator object, which we explicitly applied with the Philox algorithm. In contrast, PyTorch does not allow users to select the PRNG directly. NumPy is particularly versatile, offering robust documentation and support for a wide range of PRNGs. In our experiments with NumPy, we employed the Generator object with explicit selection of MT19937, Philox, and PCG algorithms.

Within these Python-based ML frameworks, we used high-level APIs that provide seeding functions to initialize PRNGs. For each PRNG, we generated 896 independant (randomly spaced) random streams by seeding them with 896 unique values produced using Python's built-in random number generation. It is essential for the scientific community to recognize the distinction between a seed and the full internal state of a PRNG. While the state determines the actual sequence of outputs, a seed is an input to a transformation function that initializes this state. This transformation can differ across platforms and frameworks. Because the seeding mechanism is used in all of the ML frameworks, by researcher and data scientists, we used this to initialize our random streams in our study.

For statistical quality assessment, we utilized the TestU01 BigCrush test battery, which, according to the documentation, requires slightly more than $2^{38}$ generated numbers. All data were saved in text files and processed using Jupyter Notebook, which also facilitated the execution of all bash scripts to ensure full reproducibility of the experiments.

## 4. Results and discussion

We evaluated the statistical quality of the pseudorandom numbers generated by each framework and compared these results with those obtained from the original implementations of the corresponding algorithms.

For each PRNG in each framework, we tested 896 independent random streams, each initialized with a unique, randomly spaced seed. The aggregate results are presented in Table 1.

While both PCG and Philox are advertised by their respective authors as being resistant to failure under the BigCrush test suite, our empirical evaluation suggests otherwise. Based on the 896 tested streams, we observed that these generators did not fully uphold their claimed resistance.

Nonetheless, in terms of statistical quality and reproducibility, the PCG implementation in NumPy and the Philox implementations in both TensorFlow and NumPy appear statistically consistent with their original reference implementations. In contrast, the results for MT19937 (Mersenne Twister) present certain anomalies. The reference implementation of MT is expected to consistently fail at least two tests—LinearComp tests 80 and 81. However, in our experiments, the NumPy implementation of MT fails only one of these tests consistently. Interestingly, only the PyTorch implementation of MT exhibits a statistical failure profile matching that of the original MT19937, suggesting potential deviations or platform-specific adaptations in the other implementations.

Table 1. Average BigCrush test failures per random streams, for 896 independent streams

| Generator | Total files | Avg failures per random stream |
|---|---|---|
| Pcg32 | 896 | 0.329241 |
| numpyPcg | 896 | 0.322545 |
| pytorch | 896 | 2.386161 |
| numpyMt | 896 | 1.353795 |
| Mt32 | 896 | 2.368304 |
| Tensorflow | 896 | 0.338170 |
| NumpyPhilox | 896 | 0.332589 |
| Philox32 | 896 | 0.377232 |

Across the 896 random streams tested per generator, we observed that nearly all 106 tests in the BigCrush suite were failed at least once by most implementations. Drawing from our prior study on MT19937, in which 4,096 streams were evaluated [31], we can reasonably infer that with a higher number of streams, all BigCrush tests would likely be failed at least once by each generator. This pattern suggests that such failures are not isolated anomalies but rather inherent to the statistical behavior of these PRNGs.

When comparing the overall number of distinct test failures across implementations, the results indicate little difference in statistical quality. The reference implementation of PCG32 failed 91 tests, while Philox failed 90. PyTorch's generator, which uses MT19937, also failed 91 tests, and the reference MT19937 implementation failed 95. NumPy's versions showed some variation: NumPy-PCG failed 88 tests, NumPy-MT failed 81, and NumPy-Philox failed 95. Notably, the TensorFlow implementation, which uses Philox, failed only 72 tests, performing slightly better than the others in terms of total failures.

However, while TensorFlow shows a lower overall failure count, subsequent analysis reveals that it fails certain tests not failed by the NumPy or reference implementations of Philox. This highlights the need to consider not only the number of failed tests but also which specific tests are affected, as different patterns of failure may imply different weaknesses in statistical behavior.

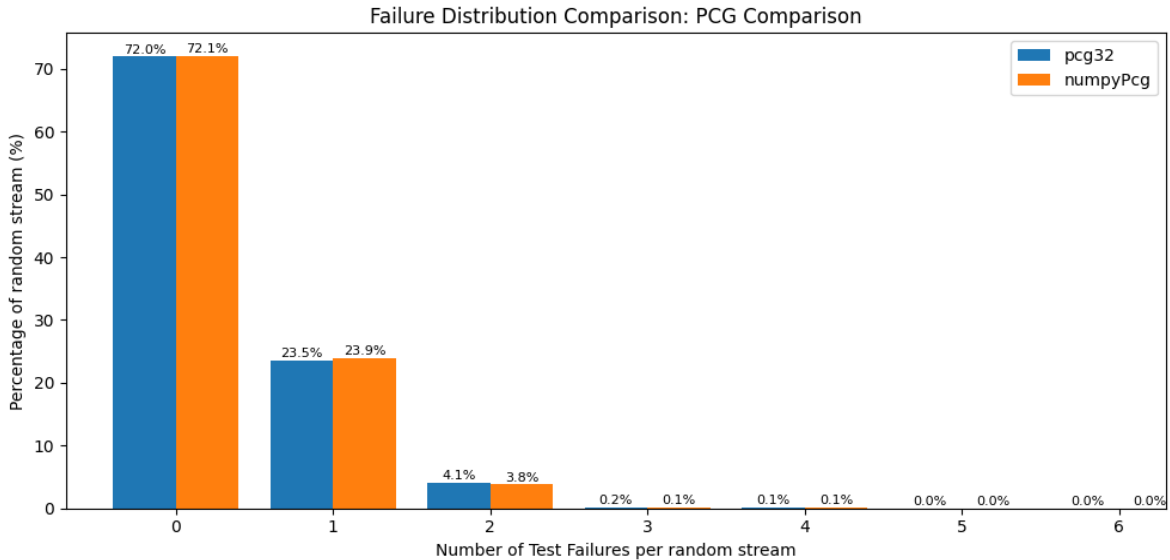

Figure 1. Number of test failures per random streams distribution, for PCG implementations

As illustrated in Figure 1, the PCG generator exhibits similar behavior in both its original implementation and the NumPy version. Notably, some individual streams of PCG fail up to four tests in the BigCrush suite—an observation not previously reported in the literature. Approximately 30% of the tested PCG streams fail at least one statistical test, which challenges the claim of "Crush resistance" made in the original PCG technical report. According to that report, PCG variants with a state size greater than 48 bits are expected to pass all BigCrush tests [3]. Our findings suggest that, in practice, a non-negligible fraction of streams may still exhibit statistical weaknesses, even when the theoretical state size requirements are met.

As shown in Figure 2, the Philox generator demonstrates similar statistical behavior across its original implementation, as well as in the TensorFlow and NumPy frameworks. We observe that certain Philox streams fail up to four tests in the BigCrush battery. Approximately 30% of the tested streams fail at least one statistical test, which stands in contrast to the claims made in the original Philox paper. Specifically, the authors assert that "*all our PRNGs pass rigorous statistical tests (including TestU01's BigCrush)*" [2]. These findings suggest that, despite its strong theoretical guarantees, Philox may still produce statistically weak streams under certain conditions in practical implementations.

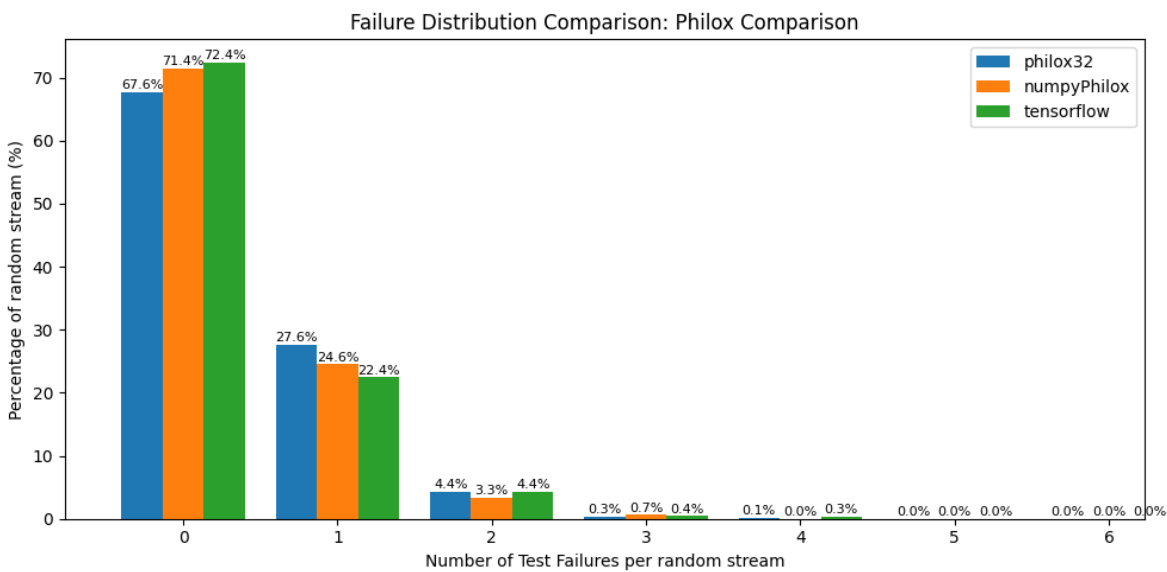

Figure 2. Number of test failures per random streams distribution, for Philox implementations

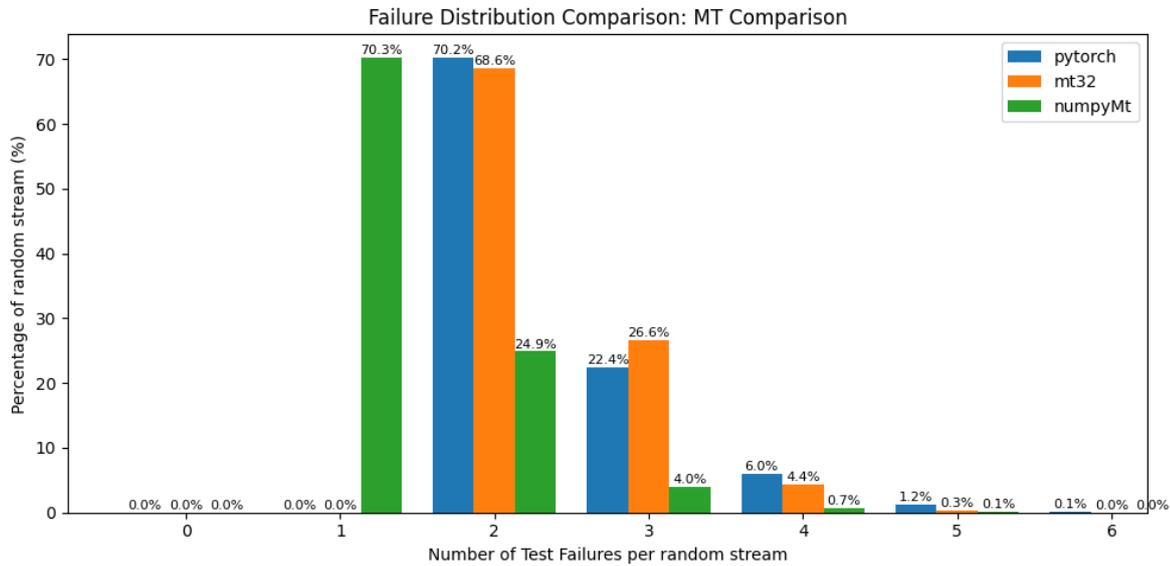

Figure 3. Number of test failures per random streams distribution, for MT implementations

Figure 3 presents more unexpected results for the MT implementations. In our previous in-depth study of MT [31], we observed that the generator consistently fails both linear complexity tests in the BigCrush suite across all random number streams. Surprisingly, the NumPy implementation of MT demonstrates better statistical performance than the original version, systematically failing only one of the two linear complexity tests. In contrast, the PyTorch implementation closely mirrors the behavior of the original MT, consistently failing both tests as expected. These differences suggest that subtle variations in implementation details can influence

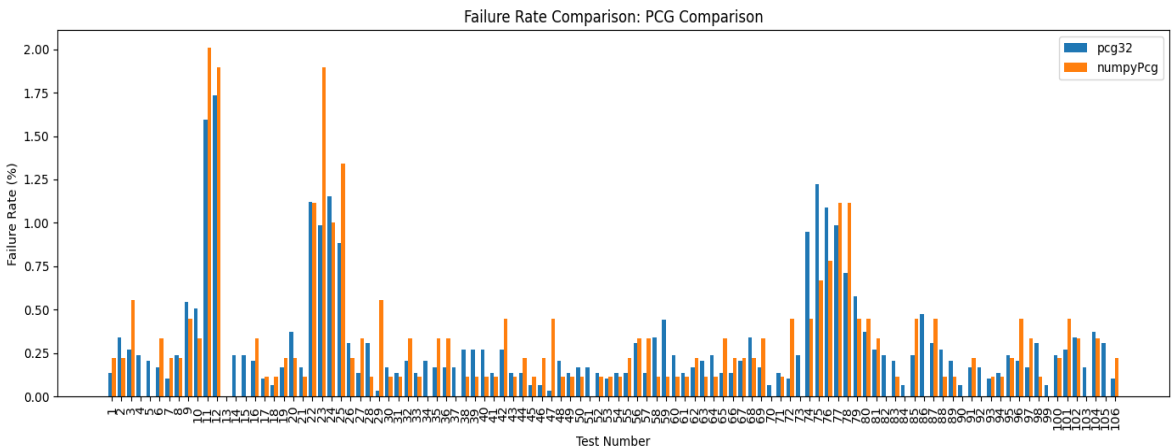

the statistical quality of the generated streams, even for well-established algorithms like MT.

Figure 4. Number of test failures per random streams distribution, for MT implementations

Figure 4 shows that the pattern of statistical test failures across all 106 BigCrush tests is highly similar between the NumPy implementation and the original reference implementation. This indicates a strong consistency in the types of weaknesses exhibited by both versions of the generator.

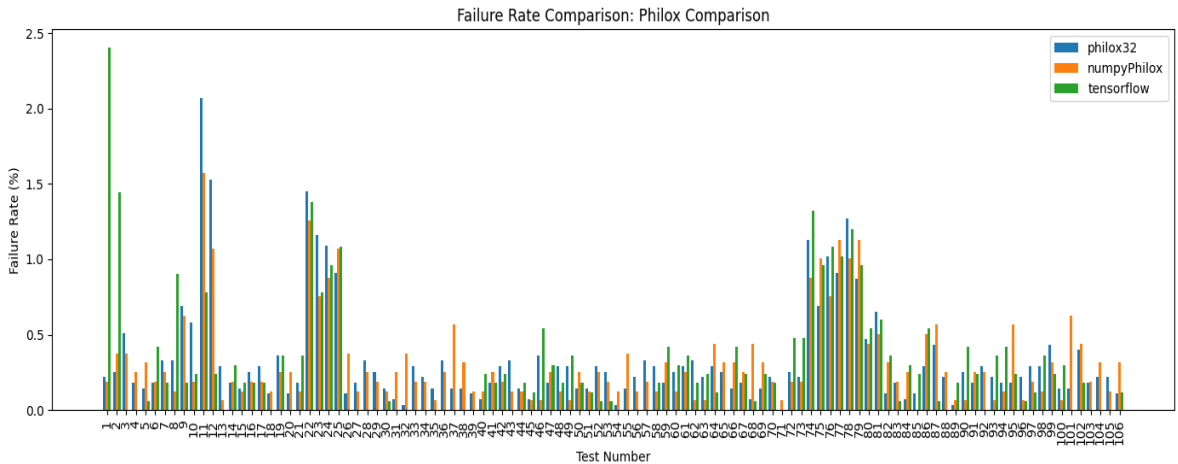

Figure 5. Number of test failures per random streams distribution, for MT implementations

Regarding the different implementations of Philox, Figure 5 reveals that the patterns of test failures are largely consistent across the original, NumPy, and TensorFlow versions. However, one notable anomaly is observed in the TensorFlow implementation, which fails statistical tests that are not failed by either the original or NumPy implementations (SerialOver). This suggests a subtle divergence in the internal behavior of TensorFlow's Philox generator that may warrant further investigation.

In Figure 6, the Y-axis has been limited to a maximum of 3% failure rate for improved visual clarity; however, it is important to note that the linear complexity tests are, in fact, failed 100% of the time. No specific anomalies or artifacts are observed in this plot, as all three Mersenne Twister implementations—PyTorch, NumPy, and the original—exhibit similar overall behavior. Nevertheless, the fact that the NumPy implementation consistently fails only one of the two linear complexity tests remains puzzling, especially given that both the original and PyTorch implementations fail both tests as expected.

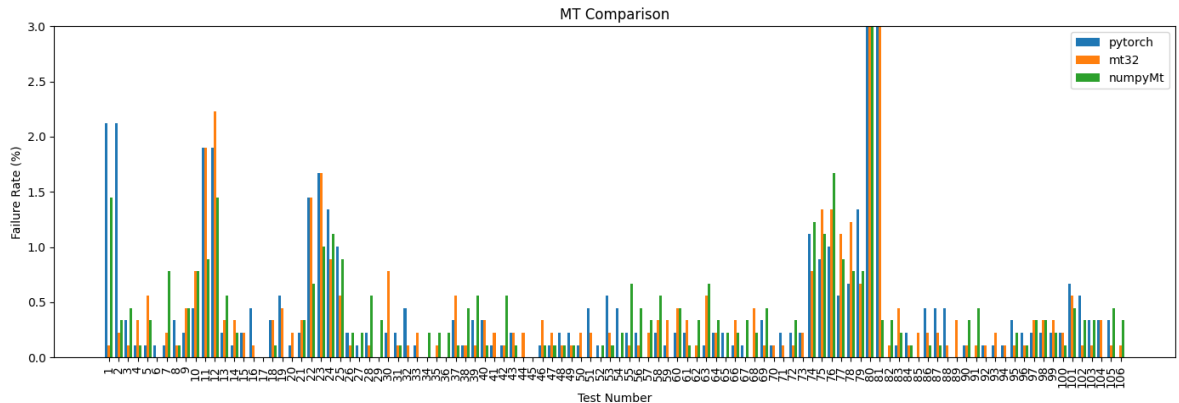

Figure 6. Number of test failures per random streams distribution, for MT implementations

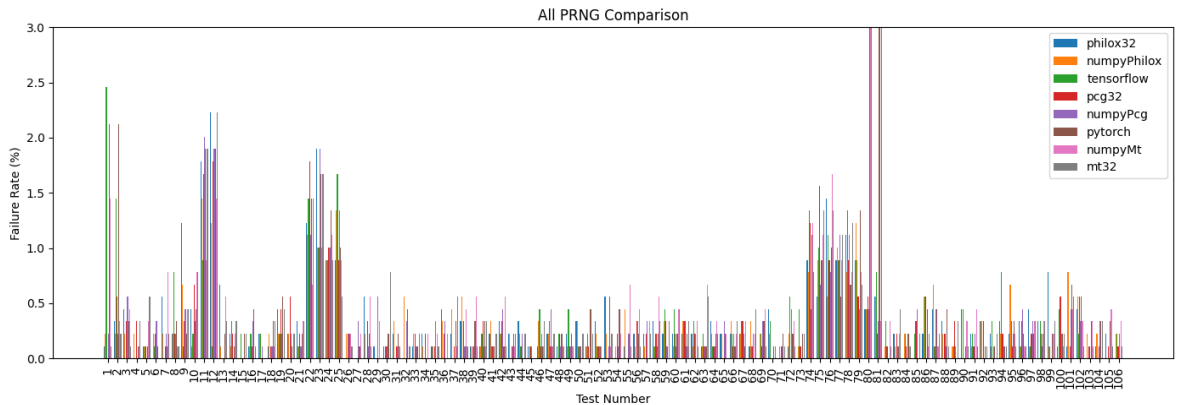

Figure 7. Number of test failures per random streams distribution, for MT implementations

As shown in Figure 7, we observe consistent overrepresentation of failures in several categories of BigCrush tests. In particular, the CollisionOver tests (tests 9–12), ClosePairs tests (22–25), and RandomWalk tests (74–79) are frequently failed across multiple generators. Additionally, the MT generator, as expected, consistently fails the Linear Complexity tests (80 and 81). An anomaly is also observed in the TensorFlow implementation, which shows increased failure rates in the SerialOver tests (1 and 2), a pattern not present in the other implementations.

In her paper [3], Melissa O'Neil describes that passing a statistical test like TestU01's BigCrush does not necessarily mean a random number generator (RNG) is of high quality. Two generators might both pass BigCrush, yet one could be significantly weaker, surviving only because the test isn't sensitive to certain flaws or because the generator has an unusually large internal state. In fact, all generators with limited memory (finite state), even ideal ones, will eventually fail statistical tests if pushed too far—they simply run out of randomness. For example, a generator with 32 bits of internal state producing 32-bit outputs can only produce each number once per cycle, which violates the natural randomness we expect, such as repeated values. This issue can be understood using the birthday paradox: in a room of just 23 people, there's already a 50% chance two share a birthday—repeats are expected in truly random sequences. When a generator's internal state is too small to allow such repeats, it starts to look suspiciously non-random. To quantify this, she defines the concept of headroom, which measures how much more state a generator has than the minimum needed to pass a given test. For instance, BigCrush is estimated to require at least 36 bits of state; a generator with 128 bits of state has 92 bits of headroom, while one that passes with just 40 bits is remarkably efficient. Estimates suggest that passing SmallCrush requires at least 32 bits, Crush 35 bits, and BigCrush 36 bits. Generators that pass these tests with minimal state are particularly noteworthy, while those needing far more may be statistically inefficient, despite passing.

From this perspective, failures in BigCrush tests by generators with state sizes below 36 bits may not indicate poor statistical quality but rather insufficient state to meet the complexity of the tests. However, all PRNGs evaluated in our study possess at least 64 bits of internal state and should, according to this reasoning, have sufficient headroom to pass BigCrush in full. The observed failures, therefore, are not due to inherently limited capacity, but likely stem from deeper statistical flaws or implementation-specific deviations that merit further scrutiny.

However, when interpreting the results of large statistical test batteries such as TestU01's BigCrush, which includes approximately 160 individual tests, it is essential to account for the effects of multiple testing. Under the null hypothesis H0, which assumes that the RNG is statistically sound, the p-values from each test should follow a uniform distribution over the interval [0,1]. Consequently, the probability of obtaining a p-value below 0.001 or above 0.999—values typically considered suspicious by BigCrush—is 0.002 for any single test. However, when multiple tests are performed independently, the likelihood of observing at least one such extreme p-value increases. Specifically, the probability of encountering at least one suspicious p-value across 160 independent tests is approximately $1-(1-0.002)^{160} \approx 27.4\%$. This implies that even for a high-quality

RNG, it is statistically expected that one or more tests may fail purely by chance. Therefore, isolated test failures should not be overinterpreted; rather, the overall pattern and consistency of failures should guide conclusions about RNG quality.

Table 2. Average BigCrush test failures per random streams, for 896 independent streams, when only considering p-values at $10^{-15}$

| Generator | Total files | Avg failures per random stream |
|---|---|---|
| Pcg32 | 896 | 0 |
| numpyPcg | 896 | 0 |
| pytorch | 896 | 2.035714 |
| numpyMt | 896 | 1.013393 |
| Mt32 | 896 | 2 |
| tensorflow | 896 | 0.039062 |
| NumpyPhilox | 896 | 0 |
| Philox32 | 896 | 0 |

In our analysis of BigCrush results, we adopted a conservative filtering approach by excluding only the most extreme p-values—those less than $10^{-15}$, reported by BigCrush as 1-eps—in order to account for potentially spurious or overly sensitive test outcomes. Using this criterion, the majority of PRNGs under evaluation, including the reference implementations of PCG32, Philox32, and their NumPy counterparts, show no statistically significant failures, supporting their classification as "Crush-resistant" under practical usage conditions. However, this context also underscores notable inconsistencies between machine learning framework implementations and their original counterparts. Specifically, the SerialOver tests (tests 1 and 2) are failed in TensorFlow, PyTorch, and NumPy-MT implementations, yet not in any of the corresponding reference generators. This points to potential anomalies introduced during integration or configuration within these frameworks. These findings emphasize the importance of not only statistical rigor but also reproducibility and implementation fidelity in machine learning libraries that rely on embedded PRNGs.

Our investigation reveals important insights into the statistical quality and reproducibility of PRNGs used within major ML frameworks. While the reference implementations of generators such as PCG32 and Philox32 exhibit strong statistical properties, often passing the entire BigCrush suite without significant anomalies, our analysis shows that their integration into ML libraries like TensorFlow, PyTorch, and NumPy can introduce reproducibility divergences in behavior.

For instance, while the SerialOver tests (tests 1 and 2) are not failed by the original implementations of Philox or MT19937, they are consistently failed by TensorFlow, PyTorch, and NumPy-MT. These failures, although limited in scope, raise important concerns about implementation fidelity and reproducibility. Similarly, all versions of MT were found to systematically fail the LinearComp tests (80 and 81), a known limitation of the algorithm, with the exception of NumPy-MT, which surprisingly fails only one of the two.

Moreover, when examining statistical behavior across large-scale replications, we find that even "Crush-resistant" generators like PCG32 and Philox can fail up to four different BigCrush tests in certain streams—a result not previously reported in the literature. In our earlier, in-depth analysis of MT19937 [31], we observed that over 4,096 replications, all 106 BigCrush tests were failed at least once. This suggests that, given a large enough sample of streams, even high-quality PRNGs may exhibit failures across the full range of tests.

Importantly, the relationship between p-value distribution and multiple testing effects must be carefully interpreted. As discussed, even a perfect RNG is expected to produce a p-value outside the [0.001, 0.999] interval in roughly 0.2% of cases. Over 160 tests, this translates to a 27.4% chance of at least one such extreme p-value occurring purely by chance. To mitigate the risk of overinterpreting such results, we conservatively excluded only extremely suspicious p-values (i.e., $p<10^{-15}$) from our summary statistics, allowing us to focus on systematic rather than incidental failures.

From a practical standpoint, these statistical deviations matter. Pseudorandom number generation plays a critical role in training neural networks, including in weight initialization, data shuffling, and stochastic optimization procedures such as stochastic gradient descent (SGD). Ensuring the quality and reproducibility of PRNG streams is thus central to the robustness and replicability of ML models. While our results suggest that ML implementations can approximate the statistical quality of their low-level C counterparts, they also reveal limitations in reproducibility and consistency across platforms and libraries.

Finally, we note that ensuring reproducible pseudorandomness in ML pipelines is non-trivial. In high-performance computing (HPC) environments, the choice of PRNG and its integration can have substantial implications not only for statistical quality but also for computational cost, energy consumption, and scalability. Future work will focus on analyzing the impact of the statistical quality of the PRNG on the quality of results of ML training. This highly important subject have not been deeply studied.

## 5. CONCLUSION

PRNGs are foundational components of ML frameworks, playing a critical role in neural network training through processes such as weight initialization, data shuffling, and optimization. While the integration of stochasticity has contributed significantly to the success of modern ML, our study highlights that the statistical properties of the underlying PRNGs remain insufficiently scrutinized —particularly in the context of their implementation in high-level frameworks.

Through comparative analysis of several PRNGs across reference C implementations and their NumPy, TensorFlow, and PyTorch counterparts, we observed that, in many cases, the statistical quality is preserved. However, certain generators exhibited unexpected test failures, even when labeled as "Crush-resistant." This raises concerns about relying solely on theoretical guarantees without empirical validation across large-scale replications. Moreover, while NumPy recommends avoiding PCG for parallel computation and instead suggests PCG64DXSM, even the latter has shown weaknesses in independent studies, such as those reported by Vigna (https://pcg.di.unimi.it/pcg.php) (https://github.com/numpy/numpy/issues/16313).

Given the increasing computational scale of ML applications and the importance of reproducibility, selecting statistically robust and implementation-safe PRNGs is becoming ever more critical. Alternative generators like xoroshiro128++ [32] offer high performance, but must also be evaluated carefully, particularly with respect to their seeding behavior and suitability for parallel execution.

Finally, although the direct impact of PRNG quality on ML model performance is not yet well understood, recent findings indicate that even subtle differences in randomness can influence training outcomes. This connection warrants further interdisciplinary research at the intersection of statistical testing, ML system design, and reproducibility standards. As ML continues to scale across domains and hardware platforms, robust, transparent, and testable PRNG selection should be treated as a foundational design choice rather than an afterthought..

**FUNDING INFORMATION**

This research received no external funding

**DATA AVAILABILITY STATEMENT**

All data are available at https://github.com/benjaminantunes/random-numbers-machine-learning

**ACKNOWLEDGEMENTS**

We acknowledge LIMOS, from Clermont Auvergne University (UCA), for the computing facilities.

**CONFLICTS OF INTEREST**

The authors declare that they have no conflicts of interest to this work

**REFERENCES**

[1] M. Matsumoto, and T. Nishimura, "Mersenne Twister: A 623-dimensionally equidistributed uniform pseudo-random number generator," *ACM Transactions on Modeling and Computer Simulation*, vol. 8, no 1, pp. 3–30, 1998.

[2] J. K. Salmon, M. A. Moraes, R. O. Dror, and D. E. Shaw, "Parallel random numbers: As easy as 1, 2, 3", in *International Conference for High Performance Computing, Networking, Storage and Analysis*, pp. 1–12, 2011.

[3] M. E. O'neill, "PCG: A family of simple fast space-efficient statistically good algorithms for random number generation," *ACM Transactions on Mathematical Software. Claremont, CA: Harvey Mudd College. https://www.cs.hmc.edu/tr/hmc-cs-2014-0905.pdf*, 2014.

[4] P. L'Ecuyer, and R. Simard, "TestU01: AC library for empirical testing of random number generators," *ACM Transactions on Mathematical Software*, vol. 33, no 4, pp. 1–40, 2007.

[5] M. Saito, and M. Matsumoto, "SIMD-oriented fast Mersenne Twister: A 128-bit pseudorandom number generator," in *Monte Carlo and Quasi-Monte Carlo Methods*, pp. 607–622, 2006.

[6] A. Rukhin, J. Soto, J. Nechvatal, M. Smid, E. Barker, S. Leigh, and A. Heckert, "A statistical test suite for random and pseudorandom number generators for cryptographic applications," *National Institute of Standards and Technology*, USA, 2001.

[7] M. Roucairol and T. Cazenave, "Comparing search algorithms on the retrosynthesis problem," *Molecular Informatics*, vol. 43, p. e202300259, 2024.

[8] C. Drummond, "Replicability is not reproducibility: Nor is it good science," in *Proc. Evaluation Methods for Machine Learning Workshop*, pp. 1–4, 2009.

[9] M. Hart, K. Idanwekhai, V. M. Alves, A. J. Miller, J. L. Dempsey, J. F. Cahoon, and A. Tropsha, "Trust not verify? The critical need for data curation standards in materials informatics," *Chemistry of Materials*, vol. 36, pp. 9046–9055, 2024.

[10] B. Antunes and D. R. C. Hill, "Reproducibility, replicability and repeatability: A survey of reproducible research with a focus on high performance computing," *Computer Science Review*, vol. 53, p. 100655, 2024.

[11] M. Huk, K. Shin, T. Kuboyama, and T. Hashimoto, "Random number generators in training of contextual neural networks," in *Asian Conf. on Intelligent Information and Database Systems*, pp. 717–730, 2021.

[12] A. Koivu, J. P. Kakko, S. Mäntyniemi, and M. Sairanen, "Quality of randomness and node dropout regularization for fitting neural networks," *Expert Systems with Applications*, vol. 207, p. 117938, 2022.

[13] Y. Lu, S. Y. Meng, and C. De Sa, "A general analysis of example-selection for stochastic gradient descent," in *International Conf. on Learning Representations*, p. 44, 2022.

[14] J. Antorán, J. Allingham, and J. M. Hernández-Lobato, "Depth uncertainty in neural networks," *Advances in Neural Information Processing Systems*, vol. 33, pp. 10620–10634, 2020.

[15] A. Mumuni and F. Mumuni, "Data augmentation: A comprehensive survey of modern approaches," *Array*, vol. 16, p. 100258, 2022.

[16] F. Maleki, K. Ovens, R. Gupta, C. Reinhold, A. Spatz, and R. Forghani, "Generalizability of machine learning models: Quantitative evaluation of three methodological pitfalls," *Radiology: Artificial Intelligence*, vol. 5, no. 1, p. e220028, 2022.

[17] I. Tsamardinos, E. Greasidou, and G. Borboudakis, "Bootstrapping the out-of-sample predictions for efficient and accurate cross-validation," *Machine Learning*, vol. 107, pp. 1895–1922, 2018.

[18] Y. Liu, S. Liu, Y. Wang, F. Lombardi, and J. Han, "A survey of stochastic computing neural networks for machine learning applications," *IEEE Trans. Neural Netw. Learn. Syst.*, vol. 32, no. 7, pp. 2809–2824, 2020.

[19] M. Magris and A. Iosifidis, "Bayesian learning for neural networks: An algorithmic survey," *Artificial Intelligence Review*, vol. 56, no. 10, pp. 11773–11823, 2023.

[20] R. Wei and A. Mahmood, "Recent advances in variational autoencoders with representation learning for biomedical informatics: A survey," *IEEE Access*, vol. 9, pp. 4939–4956, 2020.

[21] P. Ladosz, L. Weng, M. Kim, and H. Oh, "Exploration in deep reinforcement learning: A survey," *Information Fusion*, vol. 85, pp. 1–22, 2022.

[22] L. Xiao, Z. Zhang, K. Huang, and Y. Peng, "Noise optimization in artificial neural networks," in *IEEE Int. Conf. on Automation Science and Engineering*, pp. 1595–1600, 2022.

[23] K. Kim, J. Kim, J. Yu, J. Seo, J. Lee, and K. Choi, "Dynamic energy-accuracy trade-off using stochastic computing in deep neural networks," in *Annual Design Automation Conf.*, pp. 1–6, 2016.

[24] Y. Liu, Y. Wang, F. Lombardi, and J. Han, "An energy efficient online learning stochastic computational deep belief network," *IEEE Journal Emerging and Selected Topics in Circuits and Systems*, vol. 8, no. 3, pp. 454–465, 2018.

[25] S. R. Dubey and S. K. Singh, "Transformer-based generative adversarial networks in computer vision: A comprehensive survey," *IEEE Trans. Artificial Intelligence*, vol. 5, no. 10, pp. 4851–4867, 2024.

[26] D. Pranav, I. Shumailov, and R. Anderson, "Machine learning needs better randomness standards: Randomised smoothing and PRNG-based attacks," *arXiv preprint, arXiv:2306.14043*, 2023.

[27] A. Daniely and G. Vardi, "From local pseudorandom generators to hardness of learning," *in Conf. on Learning Theory*, pp. 1358–1394, 2021.

[28] J. Hu, K. Zhu, S. Cheng, N. M. Kovalchuk, A. Soulsby, M. J. H. Simmons, and R. Arcucci, "Explainable AI models for predicting drop coalescence in microfluidics device," *Chemical Engineering Journal*, vol. 481, p. 148465, 2024.

[29] K. Zhu, S. Cheng, N. Kovalchuk, M. Simmons, Y. K. Guo, O. K. Matar, and R. Arcucci, "Analyzing drop coalescence in microfluidic devices with a deep learning generative model," *Physical Chemistry Chemical Physics*, vol. 25, no. 23, pp. 15744–15755, 2023.

[30] O. E. Gundersen, K. Coakley, C. Kirkpatrick, and Y. Gil, "Sources of irreproducibility in machine learning: A review," *arXiv preprint, arXiv:2204.07610*, 2022.

[31] B. Antunes, C. Mazel, and D. R. C. Hill, "Identifying quality Mersenne Twister streams for parallel stochastic simulations," in *Winter Simulation Conf. (WSC)*, pp. 2801–2812, 2023.

[32] D. Blackman and S. Vigna, "Scrambled linear pseudorandom number generators," *ACM Trans. Math. Softw.*, vol. 47, no. 4, pp. 1–32, 2021.